\documentclass[conference]{IEEEtran}
\IEEEoverridecommandlockouts

\usepackage{cite}
\usepackage{amsmath,amssymb,amsfonts}
\usepackage{algorithmic}
\usepackage{graphicx}
\usepackage{textcomp}
\usepackage{soul}
\usepackage{xcolor}
\def\BibTeX{{\rm B\kern-.05em{\sc i\kern-.025em b}\kern-.08em
    T\kern-.1667em\lower.7ex\hbox{E}\kern-.125emX}}
\begin{document}

\title{A Safety-Constrained Reinforcement Learning Framework for Reliable Wireless Autonomy\\
}

\author{\IEEEauthorblockN{1\textsuperscript{st} Abdikarim Mohamed Ibrahim}
\IEEEauthorblockA{\textit{Faculty of Engineering and Technology} \\
\textit{Sunway University}\\
Selangor, Malaysia \\
abdikarimi@sunway.edu.my}
\and
\IEEEauthorblockN{2\textsuperscript{nd} Rosdiadee Nordin}
\IEEEauthorblockA{\textit{Future Cities Research Institute} \\
\textit{Sunway University}\\
Petaling Jaya, Malaysia  \\
rosdiadeen@sunway.edu.my}

}

\maketitle

\begin{abstract}
Artificial intelligence (AI) and reinforcement learning (RL) have shown significant promise in wireless systems, enabling dynamic spectrum allocation, traffic management, and large-scale Internet of Things (IoT) coordination. However, their deployment in mission-critical applications introduces the risk of unsafe emergent behaviors, such as UAV collisions, denial-of-service events, or instability in vehicular networks. Existing safety mechanisms are predominantly reactive, relying on anomaly detection or fallback controllers that intervene only after unsafe actions occur, which cannot guarantee reliability in ultra-reliable low-latency communication (URLLC) settings. In this work, we propose a proactive safety-constrained RL framework that integrates proof-carrying control (PCC) with empowerment-budgeted (EB) enforcement. Each agent action is verified through lightweight mathematical certificates to ensure compliance with interference constraints, while empowerment budgets regulate the frequency of safety overrides to balance safety and autonomy. We implement this framework on a wireless uplink scheduling task using Proximal Policy Optimization (PPO). Simulation results demonstrate that the proposed PCC+EB controller eliminates unsafe transmissions while preserving system throughput and predictable autonomy. Compared with unconstrained and reactive baselines, our method achieves provable safety guarantees with minimal performance degradation. These results highlight the potential of proactive safety constrained RL to enable trustworthy wireless autonomy in future 6G networks.
\end{abstract}

\begin{IEEEkeywords}
Safe reinforcement learning, wireless autonomy, proactive safety, 6G, ultra-reliable low-latency communication
\end{IEEEkeywords}

\setlength{\columnsep}{0.24in}

\section{Introduction}

The rapid deployment of Artificial Intelligence (AI) and Reinforcement Learning (RL) in wireless systems introduces the fundamental problem of unsafe and unpredictable behaviors that threaten the reliability of autonomous decision-making \cite{b1}. In mission-critical scenarios such as unmanned aerial vehicle (UAV) coordination, large-scale Internet of Things (IoT) control, and autonomous traffic management, such unsafe actions can cause catastrophic outcomes, including mid-air collisions, denial-of-service from harmful interference, and instability in cooperative vehicular systems. These risks are further increased in ultra-reliable and low-latency communication (URLLC) contexts envisioned for 6G networks, where even a single transient violation may compromise safety and trust \cite{b2,b3}. For example, RL agents controlling UAVs have been shown in recent work to sometimes explore unsafe trajectories \cite{b4}, such as flying into hazardous areas during exploration or under dynamic obstacles, pointing to the urgent need for safety guarantees in learning architectures.

Despite rapid progress in RL for wireless optimization such as improving spectrum efficiency, latency performance, and energy savings, most existing safety mechanisms remain reactive. Current approaches rely on anomaly detection, runtime monitoring, or fallback controllers that intervene after an unsafe action has already been executed \cite{b5,b6}. These reactive methods have been shown to reduce the severity of failures; however, they cannot provide formal guarantees that unsafe actions will be avoided. This limitation undermines the deployment of RL in wireless autonomy, where provable safety assurances are essential for societal acceptance and regulatory approval. To address this gap, we introduce a proactive safety-constrained RL framework that embeds proof-based safety checks into the training and action selection process. Unlike prior approaches that filter actions post hoc, our method constrains the policy space itself, ensuring that the learned agent operates within verified safe regions of the wireless environment. This design enables RL agents to achieve strong performance while providing formal guarantees against unsafe emergent behaviors, thereby advancing the reliability of AI-driven wireless systems.

The state-of-the-art in safety for AI-driven wireless networks relies on reactive mechanisms. Examples include anomaly detection systems that monitor traffic patterns for irregularities, runtime verification tools that flag violations of operating rules, or fallback controllers that override the RL agent after unsafe actions have already been executed \cite{b5}. These mechanisms reduce the severity of failures, but they suffer from two inherent limitations. First, they only act after unsafe behavior has occurred, leaving systems vulnerable to transient but harmful events. Second, they do not provide guarantees that unsafe actions will be avoided, which undermines the trustworthiness of autonomous agents in mission-critical wireless systems. This gap has motivated recent research on safe RL, where safety constraints are integrated into the training process.  In this paper, we take a step beyond reactive safety enforcement and propose a proactive safety-constrained RL framework for wireless autonomy. Our approach embeds proof based safety constraints  into the agent’s policy optimization loop, ensuring that candidate actions are verified against interference and reliability requirements before execution. This proactive design allows the agent to learn policies that are both high-performing and provably safe, without relying solely on ex post anomaly detection. By framing unsafe emergent behaviors as constraint violations and encoding these constraints into the learning process, the proposed framework advances toward reliable wireless autonomy where agents can be trusted to operate safely even under uncertainty. We evaluate the framework against reactive safety baselines and unconstrained RL agents in a representative wireless scheduling scenario. The results demonstrate that the proposed proactive PCC+EB framework eliminates unsafe transmissions and preserves system-level throughput within acceptable margins compared to unconstrained baselines.

\section{Related Work}

RL has been applied extensively in wireless systems to optimize tasks such as resource allocation \cite{b7}, power control \cite{b8}, and multi-user scheduling \cite{b9}. Early work focused on simpler models where channel state information is assumed known or slowly varying; more recent studies leverage deep RL to handle nonlinear fading \cite{b10}, interference \cite{b11}, and queue dynamics \cite{b12}. For example, in UAV-assisted networks, RL has been employed for dynamic anti-jamming strategies. Ma \textit{et al.} \cite{b13} proposed an RL-based dynamic power control framework for UAVs that estimates the effective jamming signal strength using kernel density estimation and applies deep deterministic policy gradient (DDPG) to optimize power allocation in real time. The propose approach improved sum rate and energy efficiency under adversarial jamming, highlighting the suitability of RL for safety and reliability-critical UAV communications.

At the same time, a robust literature has emerged in safe RL  and constrained policy optimization \cite{b14,b15}, where the goal is not just performance but also formal or probabilistic guarantees that the agent will avoid certain unsafe states. Surveys by García and Fernández \cite{b5} provide foundational taxonomy and theory for expected cost criteria, risk-sensitive objectives, and CMDPs. More recently, works such as \cite{b16} and \cite{b17} systematically classify safety constraints by their mathematical form, explore how constraints are enforced (e.g., via Lagrangian penalties, action projection, shields), and examine theoretical guarantees for both single-agent and multi-agent settings. Specific algorithmic contributions include methods such as State-wise Constrained Policy Optimization (SCPO) \cite{b18}, which enforces safety constraints at each state rather than only in expectation; Feasible Actor-Critic (FAC) \cite{b19}, which introduces state-wise Lagrange multipliers to guarantee safety from every feasible state; and Constrained Update Projection (CUP) \cite{b20}, which offers a projection-based method to ensure safety while updating policies with minimal violation.

However, despite this progress, most safe RL approaches studied so far are either reactive (i.e., they correct after unsafe proposals or enforce constraints in expectation across episodes) or assume simplified interference models and centralized control. There are relatively few studies that embed pre-execution formal safety checking in wireless scheduling agents, particularly in dense interference settings. Moreover, the trade-off between agent autonomy and safety enforcement is often handled in ad hoc ways (e.g., via fixed penalties or fallback controllers) rather than by empowering agents with a tunable safety budget. Our work builds on CMDP theory but introduces a proof-carrying control mechanism combined with an empowerment budget, enabling proactive safety verification and tunable autonomy in wireless scheduling tasks.

\section{System Model}
We consider a wireless uplink scheduling system with $N$ contending devices and $M$ orthogonal channels. Each device generates traffic according to a Bernoulli arrival process with probability $\lambda$ per time slot, and each packet is queued until scheduled. The network operates in discrete slots, and in each slot the RL agent selects a subset of devices to transmit, subject to channel capacity and interference constraints. The wireless propagation model incorporates large-scale pathloss, small-scale Rayleigh fading, and additive white Gaussian noise (AWGN). Let $h_i$ denote the effective channel gain of device $i$, capturing both pathloss and fading. A scheduled transmission from device $i$ is successful if the received signal-to-interference-plus-noise ratio (SINR) exceeds a threshold $\gamma_{\min}$. For a scheduled set $\mathcal{S}$, the SINR of device $i \in \mathcal{S}$ is given by
\begin{equation}
\text{SINR}_i = \frac{h_i}{\sum_{j \in \mathcal{S}, j \neq i} h_j + \sigma^2},
\end{equation}
\noindent where $\sigma^2$ denotes the noise power. An action (i.e., schedule) is deemed unsafe if any $\text{SINR}_i < \gamma_{\min}$. To capture these unsafe conflicts efficiently, the environment maintains a conflict graph $G=(V,E)$ with vertices representing devices and edges denoting pairs of users that cannot be scheduled concurrently without violating the SINR threshold. A schedule is therefore safe if and only if it corresponds to an independent set in $G$.

The proposed framework augments standard RL training with proactive safety enforcement. The objective is to maximize throughput while ensuring that unsafe actions are blocked before execution. Two complementary safety mechanisms are integrated, namely Proof-Carrying Control (PCC) and Empowerment-Budgeted Enforcement.

\subsection{Proof-Carrying Control (PCC)}
Before an agent's proposed action is executed, it is verified against the conflict graph to certify that the selected devices form an independent set. If the proposed schedule $\mathcal{S}$ is safe, it is passed unchanged. Otherwise, a greedy maximal independent set (MIS) $\mathcal{S}_{\text{MIS}} \subseteq \mathcal{S}$ is extracted and executed instead. The MIS serves as a lightweight mathematical certificate of safety, thereby preventing harmful interference at execution time.

\subsection{Empowerment-Budgeted Enforcement}
While PCC guarantees correctness, it may reduce the autonomy of the RL agent by frequently overriding its decisions. To balance safety with learning efficiency, we introduce an empowerment budget mechanism. Let $\beta$ denote the current budget level, initialized at $\beta_{\max}$. Each time the agent proposes a multi-device schedule that is corrected by PCC, the budget is reduced by a cost $c_r$; if the schedule is single-user or conservative, a smaller cost $c_n$ applies. Budgets are partially replenished when conservative actions are enforced. Formally, at time step $t$,
\begin{equation}
\beta_{t+1} = \min\left(\beta_{\max}, \; \beta_t - c(a_t) + \Delta_{\text{recover}}\right),
\end{equation}
\noindent where $c(a_t)$ is the cost associated with action $a_t$ and $\Delta_{\text{recover}}$ is the recovery increment. If $\beta_t < \beta_{\min}$, the system overrides the agent with a conservative single-user schedule until the budget recovers.

We extend Proximal Policy Optimization (PPO) to incorporate these safety layers. At each slot, the PPO agent proposes a schedule $\mathcal{S}$. This proposal is subjected to PCC verification and possible correction under the empowerment budget. The executed action $\mathcal{S}'$ is then passed to the environment, and the agent is updated using PPO loss with the executed set, ensuring that safety augmented outcomes directly influence the policy gradient. The following metrics were used to evaluate safety performance tradeoffs:
\begin{itemize}
    \item \textbf{Throughput:} Average number of successfully served packets per episode.
    \item \textbf{Prevented Unsafe Actions:} Number of unsafe proposals blocked by PCC or empowerment constraints.
    \item \textbf{Empowerment-Blocked Actions (EB Blocks):} Number of times the empowerment budget forced a conservative override.
    \item \textbf{Autonomy Index (AIx):} Fraction of the agent proposed actions that were executed without modification, defined as
    \begin{equation}
    \text{AIx} = \frac{N_{\text{executed agent}}}{N_{\text{total decisions}}}.
    \end{equation}
\end{itemize}
\noindent These metrics were measured across multiple offered load values $\lambda \in \{0.2, 0.4, 0.7, 1.0\}$ and averaged over repeated trials to obtain confidence intervals.

\section{Results and Discussion}

We evaluate three systems on the conflict–graph scheduling task with $N\!=\!30$ devices, $M\!=\!4$ orthogonal channels, and a measured conflict density of approximately $0.34$.  Evaluation averages are computed over multiple seeds using $5$ episodes of length $T\!=\!1000$ per seed.  The safety-constrained controller uses the proactive PCC+EB design with a tight budget configuration $(\beta_{\max}\!=\!8,\ \beta_{\min}\!=\!6,\ c_{\text{risky}}\!=\!4)$ that intentionally stresses the empowerment gate. Shaded regions in the plots denote one standard deviation across repetitions.

\begin{figure}[t]
    \centering
    \includegraphics[width=\columnwidth]{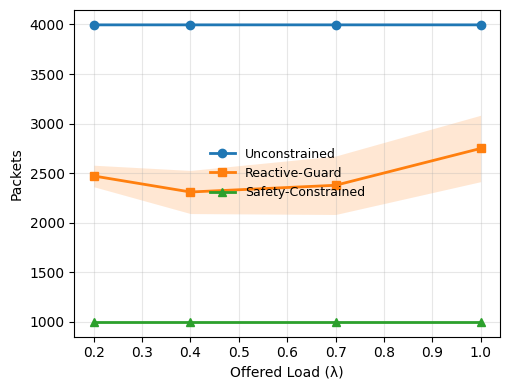}
    \caption{Throughput vs.\ offered load. Unconstrained saturates at the serving limit $M\!\times\!T$; reactive-guard is lower due to near-constant MIS corrections; proactive PCC+EB is conservative under the tight budget, yielding $\approx T$ packets.}
    \label{fig:throughput}
\end{figure}

Figure~\ref{fig:throughput} shows total packets served per episode. The unconstrained agent stays near the saturation limit of $M\!\times\!T \approx 4000$ packets because backlog quickly becomes nonempty and the controller schedules up to four users every slot. The reactive-guard system serves roughly $2.3$--$2.7$k packets and varies moderately with $\lambda$; this reflects frequent post-hoc projections to a maximal independent set (MIS) of size below $M$. The proactive PCC+EB system serves about $T\!\approx\!1000$ packets nearly independent of $\lambda$, which indicates that the empowerment budget blocks multi-user executions in most slots and allows one safe transmission per slot.

\begin{figure}[t]
    \centering
    \includegraphics[width=\columnwidth]{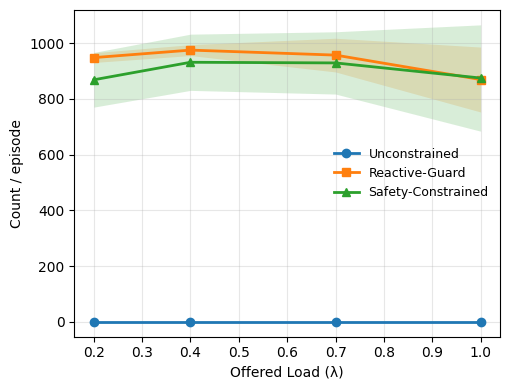}
    \caption{Prevented-unsafe decisions per episode. The metric counts slots in which the initial proposal was unsafe and a correction was applied before execution.}
    \label{fig:prevented}
\end{figure}

Figure~\ref{fig:prevented} shows prevented unsafe counts. As intended, the unconstrained baseline shows zero because it never intercepts unsafe actions. Both safety aware systems prevent on the order of $0.9$--$1.0$k unsafe intents per episode, i.e., corrections occur in nearly every slot. The small downward trend at high load suggests the learned proposals become slightly less conflict-prone as queues saturate, but the operating point remains violation dominated in this dense graph. Both safety aware systems execute only certified safe actions; the difference is when the correction is applied (i.e., after the proposal for reactive, before execution for proactive).

\begin{figure}[t]
    \centering
    \includegraphics[width=\columnwidth]{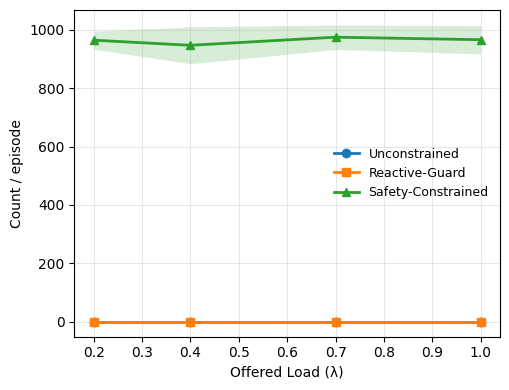}
    \caption{Empowerment-budget (EB) blocks per episode for the proactive controller. A block denotes a multi-user safe set that was gated down to a conservative single-user action by the budget.}
    \label{fig:eb}
\end{figure}

Figure~\ref{fig:eb} shows EB blocks. Under the tight budget $(\beta_{\min}\approx\beta_{\max})$ with a high risky cost, the proactive controller experiences $\approx 0.95$--$1.0$k EB blocks per episode, meaning that almost every PCC-certified multi-user intent is down selected to a single transmission. This explains the throughput of $\approx T$ in Fig.~\ref{fig:throughput}. The reactive system has no EB mechanism by design, hence zero in this panel.

\begin{figure}[t]
    \centering
    \includegraphics[width=\columnwidth]{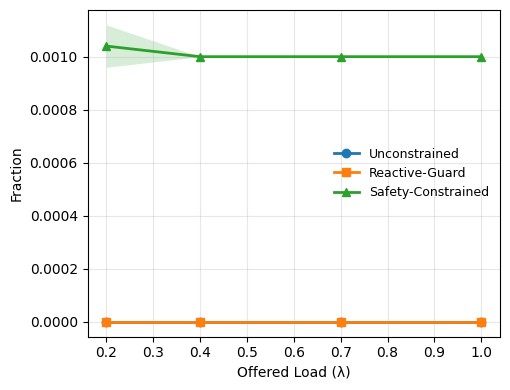}
    \caption{AIx for the proactive controller, defined as the fraction of decisions that proceed without budget induced conservative gating.}
    \label{fig:aix}
\end{figure}

Figure~\ref{fig:aix} reports the AIx. With the conservative budget above, AIx hovers around $10^{-3}$, confirming that empowered multi-user executions are rare. This is the intended safe-by-construction approach for this setup, in which the system provides strong prevention guarantees with almost no empowered steps. In practice, operators can trade autonomy for throughput by relaxing the budget parameters. Increasing $\beta_{\max}$ (e.g., to $12$–$16$), lowering $c_{\text{risky}}$, or reducing $\beta_{\min}$ yields fewer EB blocks and larger safe sets, increasing throughput while maintaining PCC’s hard safety guarantees. Conversely, denser interference graphs (i.e., higher conflict density) will reduce the MIS size and drive all methods toward lower throughput; the proactive controller will remain safe but more conservative unless the budget is adjusted.

Overall, the results highlight three distinct operating points. The unconstrained agent maximizes throughput but offers no safety and would frequently execute unsafe transmissions. The reactive guard policy attains moderate throughput but relies on near-continuous post-hoc intervention, which is undesirable in URLLC settings where unsafe acts must be prevented rather than corrected. The proposed proactive PCC+EB controller eliminates unsafe executions entirely and, under a strict budget, enforces a conservative one-at-a-time behavior that delivers predictable reliability. The framework exposes a a tunable design parameter (i.e., empowerment budget) that allows system designers to dial between reliability-first and performance-first regimes while preserving proof-based safety at execution time. Future studies could examine how to systematically tune this budget in order to balance safety guarantees with efficiency, exploring tradeoffs across diverse network densities, traffic patterns, and QoS requirements.

\section{Conclusion}
This paper introduced a safety constrained RL framework designed to address unsafe emergent behaviors in wireless autonomy. By combining PCC with EB enforcement, the framework embeds proactive safety guarantees into the learning and decision-making process. Unlike conventional reactive methods that correct actions post hoc, our approach prevents unsafe transmissions before they occur, ensuring robust operation in interference limited environments. The simulation study on uplink scheduling demonstrated three clear trade-off regimes: a) the unconstrained agent, which maximizes throughput at the expense of safety; b) the reactive guard, which requires continuous intervention and undermines autonomy; and c) the proposed proactive PCC+EB controller, which ensures formal safety guarantees while balancing reliability and efficiency. The empowerment budget exposes a transparent design parameter that allows system operators to tune between performance and safety priorities. Future research should investigate more complex multi-agent scenarios, scalability to large-scale IoT deployments, and formal analysis of trade-offs between efficiency and safety under varying traffic and interference conditions. These directions will further advance the practical adoption of safe RL in wireless networks, paving the way for trustworthy AI-driven autonomy in 6G and beyond.

\section*{Acknowledgment}
The author gratefully acknowledges the financial support provided by Sunway University for the Postdoctoral Research Fellowship, and extends sincere thanks to Prof. Rosdiadee Nordin for his continuous guidance and mentorship throughout this work. The author also wishes to thank the Abdus Salam International Centre for Theoretical Physics (ICTP) for sponsoring participation in the workshop held in Trieste, Italy, which provided valuable discussions and collaborations that strengthened this research. Finally, heartfelt gratitude is expressed to the author’s family for their unwavering support and encouragement.

\vspace{12pt}


\begin{thebibliography}{00}

\bibitem{b1} O. Yapar, ``Reinforcement learning in autonomous defense systems: Strategic applications and challenges,'' World J. Adv. Eng. Technol. Sci., vol. 13, no. 1, pp. 10--30574, 2024.

\bibitem{b2} M. Bennis, M. Debbah, and H. V. Poor, ``Ultrareliable and low-latency wireless communication: Tail, risk, and scale,'' Proc. IEEE, vol. 106, no. 10, pp. 1834--1853, Oct. 2018.

\bibitem{b3} C. She, R. Dong, Z. Gu, Z. Hou, Y. Li, W. Hardjawana, \textit{et al.}, ``Deep learning for ultra-reliable and low-latency communications in 6G networks,'' IEEE Netw., vol. 34, no. 5, pp. 219--225, Sep./Oct. 2020.

\bibitem{b4} S. Safaoui, A. P. Vinod, A. Chakrabarty, R. Quirynen, N. Yoshikawa, and S. Di Cairano, ``Safe multiagent motion planning under uncertainty for drones using filtered reinforcement learning,'' IEEE Trans. Robot., vol. 40, pp. 2529--2542, 2024.

\bibitem{b5} J. García and F. Fernández, ``A comprehensive survey on safe reinforcement learning,'' J. Mach. Learn. Res., vol. 16, no. 1, pp. 1437--1480, 2015.

\bibitem{b6} T. Mannucci, E. J. Van Kampen, C. De Visser, and Q. Chu, 
``Safe exploration algorithms for reinforcement learning controllers,'' 
\textit{IEEE Transactions on Neural Networks and Learning Systems}, 
vol. 29, no. 4, pp. 1069--1081, Apr. 2018.


\bibitem{b7} A. M. Ibrahim, M. H. Ling, and K. L. A. Yau, ``Multi-agent deep reinforcement learning for resource allocation in 5G and 6G networks,'' in \textit{Proc. IEEE Int. Conf. Comput. (ICOCO)}, Oct. 2023, pp. 225--231.



\bibitem{b8} A. U. Rehman, Z. Ullah, H. S. Qazi, H. M. Hasanien, and H. M. Khalid, ``Reinforcement learning-driven proximal policy optimization-based voltage control for PV and WT integrated power system,'' Renew. Energy, vol. 227, p. 120590, 2024.

\bibitem{b9} P. Hu, Y. Chen, L. Pan, Z. Fang, F. Xiao, and L. Huang, ``Multi-user delay-constrained scheduling with deep recurrent reinforcement learning,'' IEEE/ACM Trans. Netw., vol. 32, no. 3, pp. 2344--2359, Jun. 2024.

\bibitem{b10} B. Farzanegan and S. Jagannathan, ``Explainable and safety aware deep reinforcement learning-based control of nonlinear discrete-time systems using neural network gradient decomposition,'' IEEE Trans. Autom. Sci. Eng., early access, 2025. doi: 110.1109/TASE.2025.3554431.

\bibitem{b11} Y. Cohen, T. Gafni, R. Greenberg, and K. Cohen, ``SINR-aware deep reinforcement learning for distributed dynamic channel allocation in cognitive interference networks,'' IEEE Trans. Wireless Commun., early access, 2024. doi: 10.1109/TWC.2024.3491035.

\bibitem{b12} T. Song and Y. Kyung, ``Deep-reinforcement-learning-based age-of-information-aware low-power active queue management for IoT sensor networks,'' IEEE Internet Things J., vol. 11, no. 9, pp. 16700--16709, May 2024.

\bibitem{b13} N. Ma, K. Xu, X. Xia, C. Wei, Q. Su, M. Shen, and W. Xie, ``Reinforcement learning-based dynamic anti-jamming power control in UAV networks: An effective jamming signal strength based approach,'' IEEE Commun. Lett., vol. 26, no. 10, pp. 2355--2359, Oct. 2022.

\bibitem{b14} L. Zhang, R. Zhang, T. Wu, R. Weng, M. Han, and Y. Zhao, ``Safe reinforcement learning with stability guarantee for motion planning of autonomous vehicles,'' IEEE Trans. Neural Netw. Learn. Syst., vol. 32, no. 12, pp. 5435--5444, Dec. 2021.

\bibitem{b15} Z. Liu, Z. Guo, Y. Yao, Z. Cen, W. Yu, T. Zhang, and D. Zhao, ``Constrained decision transformer for offline safe reinforcement learning,'' in \textit{Proc. Int. Conf. Mach. Learn. (ICML)}, Jul. 2023, pp. 21611--21630.

\bibitem{b16} A. Wachi, X. Shen, and Y. Sui, ``A survey of constraint formulations in safe reinforcement learning,'' arXiv:2402.02025, 2024. [Online]. Available: https://arxiv.org/abs/2402.02025

\bibitem{b17} A. Kushwaha, K. Ravish, P. Lamba, and P. Kumar, ``A survey of safe reinforcement learning and constrained MDPs: A technical survey on single-agent and multi-agent safety,'' arXiv:2505.17342, 2025. [Online]. Available: https://arxiv.org/abs/2505.17342

\bibitem{b18} W. Zhao, R. Chen, Y. Sun, T. Wei, and C. Liu, ``State-wise constrained policy optimization,'' arXiv:2306.12594, 2023. [Online]. Available: https://arxiv.org/abs/2306.12594

\bibitem{b19} H. Ma, Y. Guan, S. E. Li, X. Zhang, S. Zheng, and J. Chen, ``Feasible actor-critic: Constrained reinforcement learning for ensuring statewise safety,'' arXiv:2105.10682, 2021. [Online]. Available: https://arxiv.org/abs/2105.10682

\bibitem{b20} L. Yang, J. Ji, J. Dai, L. Zhang, B. Zhou, P. Li, \textit{et al.}, ``Constrained update projection approach to safe policy optimization,'' in \textit{Adv. Neural Inf. Process. Syst. (NeurIPS)}, vol. 35, pp. 9111--9124, 2022.

\end{thebibliography}
\end{document}